\documentstyle[aps]{revtex}
\begin{document}
\draft
\title{one-dimensional Ising model built on small-world networks: competing dynamics}
\author{Wei Liu$^a$ Wen-Yuan Xiong$^b$and Jian-Yang Zhu$^{a,c}$\thanks{%
Corresponding author; Electronic address: zhujy@bnu.edu.cn}}
\address{$^a$Department of Physics, Beijing Normal University, Beijing\\
100875, China\\
$^b$Department of Physics, Hunan Lingling college, Yongzhou 425006, China.\\
$^c$ CCAST (World Laboratory), Box 8730, Beijing 100080, China}
\maketitle

\begin{abstract}
In this paper, we offer a competing dynamic analysis of the one-dimensional
Ising model built on the small-world network (SWN). Adding-type SWNs are
investigated in detail using a simplified Hamiltonian of mean-field nature,
and the result of rewiring-type is given because of the similarities of
these two typical networks. We study the dynamical processes with competing
Glauber mechanism and Kawasaki mechanism. The Glauber-type single-spin
transition mechanism with probability $p$ simulates the contact of the
system with a heat bath and the Kawasaki-type dynamics with probability $1-p$
simulates an external energy flux. By studying the phase diagram obtained in
the present work, we can realize some dynamical properties influenced by the
small-world effect.
\end{abstract}


Since Watts and Strogatz proposed the small-world networks (SWN) \cite{Watts}%
, which are believed to catch the essence of many network systems in nature
and society, a large number of further works have appeared (see \cite
{Albert,Strogatz,Dorogovtsev,Newman} for review). Researchers are interested
in investigating the properties of various models and processes on SWNs.
Recently, Zhu and Zhu. successfully introduced the SWN effect to the
critical dynamics of the spin system \cite{swn}, and thus extended the
investigation to the dynamic properties of spin models. In the 1960s, the
dynamic behavior of the Ising model was successfully described with the
Glauber \cite{Glauber} and Kawasaki \cite{Kawasaki} mechanisms. From then
on, great progress has been achieved based on these two mechanisms. Besides,
an interesting problem has been attracting much attention, i.e., the
competing Glauber-type and Kawasaki-type dynamics \cite
{application,Grandi,compete}. In the competing mechanism, the Glauber-type
dynamics is given probability $p$, while the Kawasaki-type one has
probability $1-p$. Zhu {\it et al.}'s recent work \cite{swn-cd} has studied
small-world network effect in the competing dynamics on the Gaussian model.
Some meaningful results has been obtained, but due to the requirement of the
convergence of the integration, they were not able to get the full phase
diagram. In this paper, we investigate the competing dynamics of the Ising
model considering the small-world network effect, and we obtain the full
phase diagram and the competing dynamic behavior. By this work, we can
further understand the influence of the SWN effect and highlight the
disparities between the dynamic mechanisms.

In the construction of SWN with a certain probability of introducing
long-range links, we will end up having a whole set of possible
realizations. Thus the theoretically correct way of treating dynamic systems
built on SWN should involve three steps: First we have to make a full list
of all the possible realizations and point out the probability of each one
of them. Second, we treat the problem independently on each network.
Finally, we give the expected value with all these results. Although being
conceptually straightforward, it is very cumbersome even for the simplest
one-dimensional Ising model. It has been suggested that the spin system on
the SWN as a whole has mean-field-like effect on individual spins due to the
long-range links\cite{Ising-1,swn}. Naturally, a simplified method of
mean-field nature was presented by Zhu {\it et al. }\cite{swn} to describe
the kinetic spin system built on SWN. According to this simplified method,
all possible networks are deemed as a single one. Then, the effective
Hamiltonian of a spin-lattice model built on such a network is defined as
the expected value over all possible realizations.

In the present work we study two specific networks: In a one-dimensional
loop, for example, (1) each randomly selected pair of vertices are
additionally connected with probability $p^A$; and (2) the vertices are
visited one after another, and its link in the clockwise sense is left in
place with probability $1-p^R$ and is reconnected to a randomly selected
other node with probability $p^R$. Networks of higher dimensions can be
built similarly. We shall refer to the first model as adding-type
small-world network ({\bf A-SWN}) and the second one as rewiring-type
network ({\bf R-SWN}).

We discuss the problem using the simplified method, and give the effective
Hamiltonian first. For the one-dimensional (1D) Ising model built on A-SWNs
and R-SWNs, the effective Hamiltonian can be written as, respectively 
\begin{equation}
-\beta {\cal H}^{A}\left( \left\{ \sigma \right\} \right)
=K\sum\limits_{k}\sigma _{k}\left[ \sigma _{k+1}+\frac{1}{2}%
p^{A}\sum\limits_{j\neq k}\sigma _{j}\right] ,  \label{H-Aswn}
\end{equation}
\begin{equation}
-\beta {\cal H}^{R}\left( \left\{ \sigma \right\} \right) =K\sum_{k}\sigma
_{k}\left[ \left( 1-p^{R}\right) \sigma _{k+1}+\frac{1}{N-1}p^{R}\sum_{j\neq
k}\sigma _{j}\right] ,  \label{H-Rswn}
\end{equation}
where $\beta =1/k_{B}T$ and $K=\beta J$. The case of $K>0$, ($J>0$),
corresponds to the ferromagnetic system.

Various dynamic processes in critical phenomena are believed to be governed
by two basic mechanisms, i.e., the Glauber-type with order parameter
nonconserved and the Kawasaki-type with order parameter conserved. Their
combination, namely the competing dynamics, gives 
\begin{equation}
\frac d{dt}P\left( \{\sigma \},t\right) =pG_{me}+\left( 1-p\right) K_{me},
\label{Comp-M}
\end{equation}
\begin{equation}
\frac d{dt}q_k(t)=pQ_k^G+(1-p)Q_k^K,\ q_i\left( t\right)
=\sum\limits_{\{\sigma \}}\sigma _iP\left( \{\sigma \},t\right) .
\label{Comp-q}
\end{equation}
This dynamic competition has also been receiving attention\cite
{compete,swn-cd}. Here, $pG_{me}$ (or $pQ_k^G$) denotes the Glauber-type
mechanism with probability $p$ and $\left( 1-p\right) K_{me}$ [or $%
(1-p)Q_k^K $] denotes the Kawasaki-type mechanism with probability $1-p$;
they are determined, respectively, by the Glauber-type single-spin
transition probability $W_i\left( \sigma _i\rightarrow \hat{\sigma}_i\right) 
$ \cite{zhu1} and by the Kawasaki-type pair-spin redistribution probability $%
W_{jl}(\sigma _j\sigma _l\rightarrow \hat{\sigma}_j\hat{\sigma}_l)$ \cite
{zhu2}.

In their original form, the Glauber-type dynamics and the Kawasaki-type
dynamics both favor a lower energy state. However, the competing dynamics is
usually used to describe a system in contact with a heat bath while exposed
to an external energy flux. Naturally one requires a competition between one
process favoring lower system energy and the other one favoring higher
system energy. Usually, the Glauber-type mechanism is used to simulate the
contact of the system with a heat bath and it favors a lower energy state.
On the other hand the Kawasaki-type mechanism can be modified to simulate an
external energy flux that drives the system towards a higher energy state.
This can be achieved by switching $\beta $ to $-\beta $, or $K=\beta
J=J/K_{B}T$ to $-K$, and modifying the redistribution probability
accordingly. This means that the competition between the Glauber-type
mechanism and the Kawasaki-type mechanism is actually a competition between
ferromagnetism and antiferromagnetism. Probing the competing behavior of the
1D Ising on A-SWNs and R-SWNs is certainly a problem of interest. The
equation of evolution of the local magnetization is given by 
\begin{equation}
\frac{d}{dt}q_{k}(t)=pQ_{k}^{G^{A,R}}+\left( 1-p\right) Q_{k}^{K^{A,R}}.
\label{dqdt}
\end{equation}
The first and the second term correspond, respectively, to the part of the
Glauber's dynamics with probability $p$, and the part of the modified
Kawasaki's dynamics (only switching $K=\beta J$ to $-K$) with probability $%
1-p$. Our calculation will focus on {\bf A-SWN}, while the result of {\bf %
R-SWN} will be given straightforwardly.

For the Glauber-type dynamics 
\begin{equation}
Q_i^{G^A}\equiv -q_i\left( t\right) +\sum\limits_{\{\sigma \}}\left[
\sum\limits_{\hat{\sigma}_i}\hat{\sigma}_iW_i^A\left( \sigma _i\rightarrow 
\hat{\sigma}_i\right) \right] P\left( \{\sigma \},t\right) ,  \label{q-G-A}
\end{equation}
and for the Kawasaki-type dynamics 
\begin{eqnarray}
Q_k^{K^A} &\equiv &-2q_k(t)-p^A(N-1)q_k(t)+\sum\limits_{\{\sigma \}}\left\{
\sum_{\omega =\pm 1}\left[ \sum\limits_{\hat{\sigma}_k,\hat{\sigma}%
_{k+\omega }}\hat{\sigma}_kW_{k,k+\omega }^A\left( \sigma _k\sigma
_{k+\omega }\rightarrow \hat{\sigma}_k\hat{\sigma}_{k+\omega }\right)
\right] \right.  \nonumber \\
&&\left. +p^A\left[ \sum\limits_{l\neq k}\sum\limits_{\hat{\sigma}_k,\hat{%
\sigma}_l}\hat{\sigma}_kW_{kl}^A\left( \sigma _k\sigma _l\rightarrow \hat{%
\sigma}_k\hat{\sigma}_l\right) \right] \right\} P(\{\sigma \},t),
\label{q-K-A}
\end{eqnarray}
where the expressions of $W_i\left( \sigma _i\rightarrow \hat{\sigma}%
_i\right) $ and $W_{jl}(\sigma _j\sigma _l\rightarrow \hat{\sigma}_j\hat{%
\sigma}_l)$ can be found in Ref. \cite{swn} (but differently from that
paper, we should switch $\beta $ to $-\beta $ for $W_{jl}$). Three important
combined terms in Eqs. (\ref{q-G-A}) and (\ref{q-K-A}) are calculated, for
the 1D Ising model, to be 
\[
\sum_{\hat{\sigma}_k}\hat{\sigma}_kW_k^A\left( \sigma _k\rightarrow \hat{%
\sigma}_k\right) =\tanh \left[ K\left( \sigma _{k-1}+\sigma _{k+1}\right)
+K\left( N-1\right) p^A\bar{\sigma}\right] , 
\]
\begin{eqnarray*}
&&\sum\limits_{\hat{\sigma}_k,\hat{\sigma}_{k\pm 1}}\hat{\sigma}_kW_{k,k\pm
1}^A\left( \sigma _k\sigma _{k\pm 1}\rightarrow \hat{\sigma}_k\hat{\sigma}%
_{k\pm 1}\right) \\
&=&\frac{\sigma _k+\sigma _{k\pm 1}}2+\frac 14\left( \sigma _{k\mp 1}-\sigma
_{k\pm 2}-\sigma _{k\mp 1}\sigma _k\sigma _{k\pm 1}+\sigma _k\sigma _{k\pm
1}\sigma _{k\pm 2}\right) \tanh \left( 2K\right) ,
\end{eqnarray*}
\begin{eqnarray*}
&&\sum\limits_{l\left( l\neq k\right) }\sum\limits_{\hat{\sigma}_k,\hat{%
\sigma}_l}\hat{\sigma}_kW_{kl}^A\left( \sigma _k\sigma _l\rightarrow \hat{%
\sigma}_k\hat{\sigma}_l\right) \\
&=&\frac 12\left( N-1\right) \left( \sigma _k+M\right) +\sum\limits_{l\left(
l\neq k\right) }\frac 12\left( \sigma _k-\sigma _l\right) \tanh \left\{
K\left[ \sigma _k\left( \sigma _{k-1}+\sigma _{k+1}-\sigma _{l-1}-\sigma
_{l+1}\right) \right] \right\} .
\end{eqnarray*}

Now we turn to determine the system behavior by studying the tendency of
evolution of the following order parameters: 
\begin{equation}
M(t)\equiv \frac 1N\sum\limits_kq_k(t),~M^{\prime }(t)\equiv \frac 1N%
\sum\limits_k(-1)^kq_k(t).  \label{Mag}
\end{equation}
Obviously, a state with both vanishing $M(t)$ and $M^{\prime }(t)$
corresponds to the disordered paramagnetic phase; a state with nonvanishing $%
M(t)$ and vanishing $M^{\prime }(t)$ corresponds to the ferromagnetic phase;
and, in an antiferromagnetic phase we will find vanishing $M(t)$ and
nonvanishing $M^{\prime }(t)$. If both order parameters are nonvanishing,
this phase cannot be simply identified as a ferromagnetic or paramagnetic
phase, but can be tentatively named as a heterophase.

The cases of A-SWN will be discussed in detail in the following, and we will
determine the tendency of system evolution under a small perturbation $%
M(t)\rightarrow 0$. We then give the results of the case of R-SWN.

For the 1D Ising model built on A-SWNs, the effective Hamiltonian of the
system is given by Eq. (\ref{H-Aswn}). From Eqs. (\ref{q-G-A})-(\ref{Mag}),
we can easily obtain 
\begin{equation}
M(t)=M_0\exp \left\{ -p\left[ \left( 1-\tanh 2K\right) -K\left( N-1\right)
p^A\left( 1-\frac 12\tanh ^22K\right) \right] t\right\} ,  \label{M-A}
\end{equation}
\begin{eqnarray}
M^{\prime }\left( t\right) &=&M_0^{\prime }\exp \left[ -\left\{ p\left(
1+\tanh 2K\right) \right. \right.  \nonumber \\
&&\left. \left. +\left( 1-p\right) \left[ 2-\tanh 2K+\frac 12\left(
N-1\right) p^A\left( 1-\frac 12\tanh 2K-\frac 14\tanh 4K\right) \right]
\right\} t\right] .  \label{M'-A}
\end{eqnarray}
The tendency of evolution of the order parameters $M(t)$ and $M^{\prime }(t)$
can be demonstrated by Eqs. (\ref{M-A}) and (\ref{M'-A}) when the system
undergoes a small perturbation.

(1) When $p^{A}=0$, it means that no long-range link exists, and 
\begin{equation}
M(t)=M_{0}\exp \left[ -p\left( 1-\frac{\tanh 2K}{\tanh 2K_{c}^{0}}\right)
t\right] ,  \label{M-regular}
\end{equation}
\begin{equation}
M^{\prime }(t)=M_{0}^{\prime }\exp \left\{ -\left[ p\left( 1+\frac{\tanh 2K}{%
\tanh 2K_{c}^{0}}\right) +\left( 1-p\right) \left( 2-\frac{\tanh 2K}{\tanh
2K_{c}^{0}}\right) \right] t\right\} ,  \label{M'-regular}
\end{equation}
where $\tanh 2K_{c}^{0}=1$ corresponds to the critical point of the
one-dimensional (1D) Ising model without the SWN effect ($%
K_{c}^{0}\rightarrow \infty ,$ $T_{c}^{0}=0$). When $K<K_{c}^{0}$, ($%
T>T_{c}^{0}=0$), Eqs. (\ref{M-regular}) and (\ref{M'-regular}) show that $%
M(t)$ and $M^{\prime }(t)$ are both approaching zero exponentially due to
the fact that $\tanh 2K<1$, and thus we can reach the conclusion that, by
whatever amount one increases the energy flux, the system will stay in the
paramagnetic phase at arbitrary finite temperature. When $K\rightarrow
K_{c}^{0}$, $M^{\prime }(t)\rightarrow 0$, and 
\[
M(t)=M_{0}\exp \left[ -p\frac{t}{\tau }\right] ,~\tau =\frac{1}{1-\tanh
\left( 2K\right) /\tanh \left( 2K_{c}^{0}\right) }\rightarrow \infty , 
\]
the critical slowing down of the order parameter $M(t)$ will appear at the
critical point $K_{c}^{0}$.

(2) Now a small portion of adding-type long links are introduced to the
system, the system behavior in this case can be described by Eqs. (\ref{M-A}%
) and (\ref{M'-A}). Obviously, $M^{\prime }(t)$ is approaching zero
exponentially at any temperature and any $p^A$. From Eqs. (\ref{M'-A}) and (%
\ref{M'-regular}), we can see clearly that the rate of $M^{\prime }(t)$
approaching zero is faster than that in the regular network. Different from $%
M^{\prime }\left( t\right) $, the evolution tendency of the order parameter $%
M\left( t\right) $ depends on both $K$ and $p^A$. We can obtained the
critical point by the following equation: 
\[
\tanh 2K_c^A+\left( N-1\right) p^AK_c^A\left( 1-\frac 12\tanh
^22K_c^A\right) =1.
\]
To give an example, if we suppose $p^A=1/N$, we can get $\left. \tanh
2K_c^A\right| _{p^A\sim 1/N}=0.6809$, or $\left. K_c^A\right| _{p^A\sim
1/N}=0.154$. Relative to $K_c^A$, when $t\rightarrow \infty $, $%
M(t)\rightarrow 0$ for $K<K_c^A$, $M(t)\neq 0$ for $K>K_c^A$, and $M(t)$
experiences critical slowing down for $K\rightarrow K_c^A$. So, combining $%
M\left( t\right) $ and $M^{\prime }(t)$ we can conclude that: (2.a) For the
case of $K<K_c^A$, the system stays in the paramagnetic phase; (2.b) For the
case of $K\rightarrow K_c^A$, the system shows the critical slowing down;
and (2.c) For the case of $K>K_c^A$, the system stays in the ferromagnetic
phase.

However, when $p=0$, we cannot identify the system simply as ferromagnetic
or paramagnetic. Because in this case, it depends on the initial state. If $%
M_0\neq 0$, the system will stay in ferromagnetic, otherwise the system will
be paramagnetic.

The phase diagram is shown in Fig. 1(a).

For the 1D Ising model built on R-SWNs, the effective Hamiltonian of the
system is given by Eq. (\ref{H-Rswn}). With analogous calculation, one can
get the equation of the critical point $K_c^R$%
\begin{equation}
\tanh \left[ 2K_c^R\left( 1-p^R\right) \right] +2K_c^Rp^R\left\{ 1-\frac 12%
\tanh ^2\left[ 2K_c^R\left( 1-p^R\right) \right] \right\} =1,
\label{Ising-Kc-R}
\end{equation}
and the time-evolution of the orders parameters $M(t)$ and $M^{\prime }(t)$%
\begin{equation}
M(t)=M_0\exp \left\{ -p\left( 1-\tanh \left[ 2K\left( 1-p^R\right) \right]
-2Kp^R\left\{ 1-\frac 12\tanh ^2\left[ 2K\left( 1-p_R\right) \right]
\right\} \right) t\right\} ,  \label{M-R}
\end{equation}
\begin{eqnarray}
M^{\prime }\left( t\right) &=&M_0^{\prime }\exp \left[ -\left\{ p\left(
1+\tanh \left[ 2K\left( 1-p_R\right) \right] \right) +\left( 1-p\right)
\left[ (1-p^R)\left\{ 2-\tanh \left[ 2K\left( 1-p_R\right) \right] \right\}
\right. \right. \right.  \nonumber \\
&&\left. \left. \left. +\frac 12p^R\left\{ 1-\frac 12\tanh \left[ 2K\left(
1-p_R\right) \right] -\frac 14\tanh \left[ 4K\left( 1-p_R\right) \right]
\right\} \right] \right\} t\right] .  \label{M'-R}
\end{eqnarray}

(1) When no rewiring long link exists, i.e., $p^R=0$, the evolution is the
same as the case of $p^A=0$.

(2) When a small portion of rewiring-type long links is introduced to the
system, $M^{\prime }(t)$ is approaching zero exponentially at any
temperature and any $p^R$. The decay rate of $M^{\prime }(t)$ is also faster
then before. Different from $M^{\prime }\left( t\right) $, the evolution
tendency of the order parameter $M\left( t\right) $ depends on both $K$ and $%
p^R$. The calculation of the critical point is similar to the A-SWN one.
When $p^R=0.1$, we can get the critical point, $\left. \tanh \left[
2K_c^R\left( 1-p^R\right) \right] \right| _{p^R=0.1}=0.90222$, or $\left.
K_c^R\right| _{p^R=0.1}=0.82446$. Relative to $K_c^R$, when $t\rightarrow
\infty $, $M(t)\rightarrow 0$ for $K<K_c^R$, $M(t)\neq 0$ for $K>K_c^R$, and 
$M(t)$ shows critical slowing down for $K\rightarrow K_c^R$. Combining $%
M\left( t\right) $ and $M^{\prime }(t)$ we can conclude that: (2.a) For the
case of $K<K_c^R$, the system stays in the paramagnetic phase; (2.b) For the
case of $K\rightarrow K_c^R$, the system shows the critical slowing down;
and (2.c) For the case of $K>K_c^R$, the system stays in the ferromagnetic
phase.

The phase diagram is shown in Fig. 1(b).

In this paper, we analytically study the dynamic properties of the 1D Ising
model built on small-world networks. Two typical SWNs are investigated, the
adding type and rewiring type. As is generally known, the 1D Ising model on
the regular lattice does not show continuous phase transition at any nonzero
temperature. However, if the SWN effect is introduced, critical phenomena
appear in the 1D Ising model. With competing dynamics, as long as $p\neq 0$,
the phase diagrams are separated into two regions. Below the critical
temperature, the system will get into the ferromagnetic phase, while above
the temperature, the system will get into the paramagnetic phase. The
critical temperature is independent of the competing probability $p$.
Different from the 2D Ising model \cite{application} and the Gaussian model 
\cite{swn-cd}, the 1D Ising model built on SWNs does not show
antiferromagentic phase at any temperature and any competing probability $p$.

As we have seen above, the Ising model shows critical phenomena on both
A-SWNs and R-SWNs. This is because random links introduce long-range
interactions. It is reasonable that the system will exhibit long-range order
at finite temperature. Furthermore, the more extra links, the higher the
critical temperature. Thus on R-SWNs the Ising model has a maximum critical
temperature for there is a maximum number of the random links (it is also
expected that the simplified method with an effective Hamiltonian will fail
when most regular links are rewired). However, long-range interactions do
not lead to antiferromagnetic order at any competition probability, but
instead, the long range links make any antiferromagnetic order decay faster.
This is because the long-range interaction here is random, while
antiferromagnet needs ordered long interactions.

The work was supported by the National Natural Science Foundation of China
(No. 10375008), and the National Basic Research Program of China
(2003CB716302).

\null\vskip0.2cm

\centerline{\bf Caption of figures} \vskip1cm

Fig.1. The phase diagrams of $1D$ Ising model built on SWN with competing
dynamics, Glauber-type with probability $p$, and Kawasaki-type with
probability ($1-p$). (a) On the A-SWN structure: $p^{A}\sim 1/N$; (b) On the
R-SWN structure: $p^{R}=0.1$. In which, $\tanh 2K_{c}^{0}=1$ corresponding
to the critical point of the $1D$ Ising model without the SWN effect ($%
K_{c}^{0}\rightarrow \infty ,$ $T_{c}^{0}=0$).

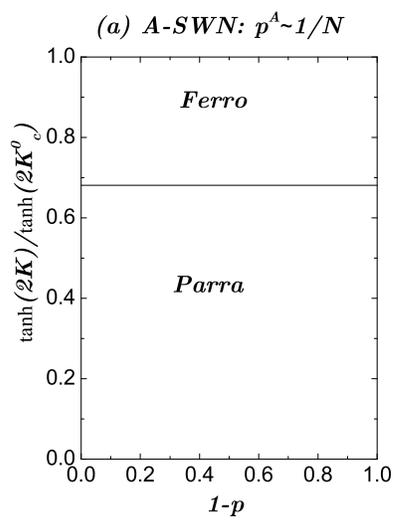
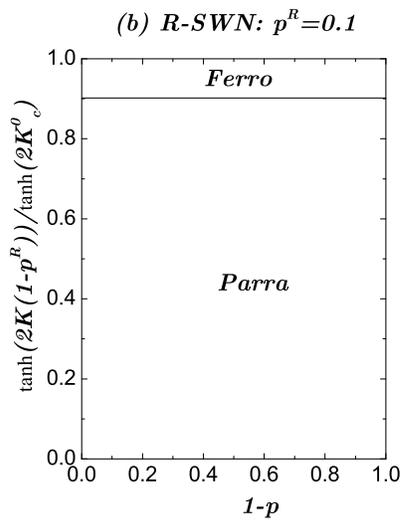
\end{document}